\documentclass[a4paper,10pt,twoside]{cpc-hepnp}
\usepackage{multirow}
\usepackage{multicol}
\usepackage{graphicx}
\usepackage{booktabs}
\usepackage{amssymb,bm,mathrsfs,bbm,amscd}
\usepackage[tbtags]{amsmath}
\usepackage{lastpage}
\usepackage[colorlinks=true,linkcolor=red,citecolor=blue]{hyperref}
\begin{document}
\fancyhead[c]{\small Chinese Physics C~~~Vol. XX, No. X (201X)XXXXXX} \fancyfoot[C]{\small 010201-\thepage}
\footnotetext[0]{Received XX March 2015}
\title{Pure Annihilation Type $B \to K_0^{*\pm}(1430)K^{(*)\mp}$ Decays in the  Family Non-universal $Z^\prime$ Model\thanks{Supported by National Natural Science Foundation of China (11175151, 11375208 and 11235005) and the Program for New Century Excellent Talents in University (NCET) by Ministry of Education of P. R. China (NCET-13-0991)}}
\author{Ying Li$^{1,3;1)}$\email{liying@ytu.edu.cn}\quad Dan-Dan Wang$^{1}$
\quad Cai-Dian Lu$^{2,3;2)}$\email{lucd@ihep.ac.cn}}
\maketitle

\address{
$^1$ Department of Physics, Yantai University, Yantai 264-005, China\\
$^2$ Institute of High Energy Physics, Chinese Academy of Sciences, Beijing 100049, China\\
$^3$ State Key Laboratory of Theoretical Physics, Institute of Theoretical Physics,\\
Chinese Academy of Sciences, Beijing 100190, China}

\begin{abstract}
By assuming that the scalar meson $K_0^*(1430)$ belongs to the first excited states or the lowest lying ground states, we study the pure annihilation-type decays $B \to K_0^{*\pm}(1430)K^{(*)\mp}$ in the QCD factorization approach. Within the standard model, the branching fractions are  at the order of $10^{-8}-10^{-7}$, which is possible to be measured in the ongoing LHCb experiment or  forthcoming Belle-II experiment.   We also study these decays in  the family non-universal $Z^\prime$ model. The results show that if $m_{Z^\prime}\approx 600\mathrm{GeV}$ ($\zeta=0.02$), both the branching fractions and $CP$ asymmetries of $\overline B^0\to K_0^{*+}(1430)K^-$ could be changed remarkably, which provides us  a place for probing the effect of new physics. These results could be used to constrain the parameters of $Z^\prime$ model.
\end{abstract}

\begin{multicols}{2}

\section{Introduction}\label{sec:1}
Heavy flavor physics has been a hot topic for many years in particle physics, since it is important for the study of CP violation   and also a good place for searching of new physics signals. With more and more data from the LHCb experiment, many hadronic B decay modes are well studied experimentally, but not the case for the theoretical side. Among them, rare decays with flavor changing neutral currents are the most interesting, since they have a quite small branching ratios in the standard model, thus more sensitive to any new physics contributions.

In this work, we shall study the impact of a family non-universal leptophobic $Z^\prime$ boson on $B \to  K_0^{*\pm}(1430)K^{(*)\mp}$ decays dominated by flavor changing neutral currents (FCNC). In the standard model, the FCNC processes are suppressed    since it is only induced by loop diagrams. The branching fractions of these decays are predicted at the order of $10^{-8}-10^{-7}$ \cite{prd}, which is possible to be measured in the ongoing LHCb experiment or  forthcoming Belle-II experiment. While the family non-universal $Z^\prime$ boson leads to the tree-level flavor changing neutral currents, which may enhance the branching ratios of these decays. Motivated by this, many FCNC processes induced by $Z^\prime$ in flavor physics have been explored extensively in literatures \cite{Barger:2004hn,changkllzp,liy}. Furthermore, as   the scalar meson $K_0^{*\pm}(1430)$ is now unclear whether it belongs to the first excited states (Scenario 1) or the ground states (Scenario 2) \cite{Cheng:2005nb}, we have to discuss both  cases in the current work.

 Unlike the leptonic decays, the non-leptonic $B$ decays are complicated because many degrees of freedom  and scales are involved. Based on the effective field theories, there are three major QCD-inspired approaches for dealing with them, namely, the QCD factorization (QCDF) \cite{QCDF}, perturbative QCD (PQCD) \cite{PQCD}, and soft collinear effective theory \cite{SCET}. In this work, we shall employ the QCDF for evaluating the relevant hadronic matrix elements, as it is a systematic framework to calculate these matrix elements from QCD theory, which holds in the heavy quark limit $m_b \to \infty$.

\section{The family non-universal Z' model}
The recent discovery  at the LHC \cite{higgs} of a light Higgs with mass around 125 GeV  opened a new window to physics beyond the Standard Model (SM). In some new physics (NP) models \cite{Langacker:2008yv}, when the initial group breaks down to $SU(2)_L\times U(1)_Y$ of SM, an extra group $U(1)^\prime$ will usually be produced, which leads to an  additional massive neutral gauge boson called $Z^\prime$. If such a gauge boson were observed, it would be a concrete evidence of the existence of NP. Based on the assumption that the $Z^\prime$ shares similar characters with $Z$ boson of SM, many efforts have been made to search for $Z^\prime$ directly by analyzing the data of dilepton. Now, at the LHC, the lower mass limit is $2.86~\mathrm{TeV}$ ($1.90~\mathrm{TeV}$) at the $95\%$ confidence level at $8~\mathrm{TeV}$ colliding energy with an integrated luminosity of $19.5 \mathrm{fb}^{-1}$ by using $e^+e^-$ and $\mu^+\mu^-$ \cite{ATLAS:2013jma} (or $\tau^+\tau^-$ \cite{ATLA:2013yha}) events. However, such constraints from LHC are invalid if the $Z^\prime$ boson does not couple or couples very weakly with the leptons, thus one has to resort to hadronic channels. Although the couplings between quarks and $Z^\prime$ are family universal in most models, the family non-universal $Z^\prime$ can also be realized in some models. For instance, such a family non-universal leptophobic $Z^\prime$ boson can be realized in $E_6$ model \cite{e6}. The phenomenological studies of family non-universal $Z^\prime$ on possible colliders have been explored in detail recently \cite{Chiang:2014yva}.

On the gauge interaction basis, the interactions of the leptophobic $Z^\prime$ boson with SM quarks can be expressed as
\begin{eqnarray}\label{zprimed}
{\cal L}^{Z^\prime}=-g_2 Z^{\prime \mu }\sum_{i,j} {\overline \psi_i^I} \gamma_{\mu} \left[ (\epsilon_{\psi_L})_{ij} P_L + (\epsilon_{\psi_R})_{ij} P_R \right] \psi^I_j ,
\end{eqnarray}
where the field $ \psi_i$ is the $i$th family fermion and $P_{L,R} = (1 \mp \gamma_5) / 2$ is the chirality projection operators. $\epsilon_{\psi_L}$ ($\epsilon_{\psi_R}$) stand for  the left-handed (right-handed) chiral couplings, and they are required to be hermitian because the Lagrangian is real. When the weak eigenstates being rotated to the physical basis, the mass eigenstates will be obtained by $\psi_{L,R}=V_{\psi_{L,R}} \psi_{L,R}^I$. Correspondingly,  the coupling matrices of down-type quarks read
\begin{eqnarray}
B^L_d  \equiv  V_{d_L} \epsilon_{d_L} V_{d_L}^{\dagger},\,\,\,\,
B^R_d \equiv  V_{d_R} \epsilon_{d_R} V_{d_R}^{\dagger}.
\end{eqnarray}
Since we do not need the couplings for up-type quarks, we will not discuss them here. Obviously, if the matrixes  $\epsilon_{d_{L,R}}$ are not proportional to the identity matrix, the nonzero off-diagonal elements in the $B^{L,R}_{u,d}$ appear, which will induce the FCNC interactions at the tree level. For simplicity, the right-handed couplings are often supposed to be flavor-diagonal. Then, the effective Hamiltonian of the $\bar b\to\bar d q \bar q $ transitions mediated by the $Z^\prime$ is given by
\begin{eqnarray}
{\cal H}_{\rm eff}^{Z^\prime}
&=&\frac{2 G_F}{\sqrt{2}} \left(\frac{g_2 m_Z}{g_1 m_{Z^\prime}}\right)^2
B^{L*}_{bd}({\bar b}d)_{V-A} \sum_q \left( B^L_{qq} ({\bar q}q)_{V-A} \right. \nonumber \\
&& \left.+ B^R_{qq} ({\bar q}q)_{V+A} \right) + \mbox{h.c.} ~,
\label{eqn:Heff1}
\end{eqnarray}
where $g_1=e/(\sin{\theta_W}\cos{\theta_W})$ and $m_{Z^{\prime}}$ denotes the mass of $Z^\prime$ boson. The diagonal elements $B_{qq}^{L,R}$ are real due to the hermiticity of the effective Hamiltonian. In contrast, the off-diagonal element $B_{bd}^{L}$ might be a complex number with a new weak phase $\phi_{bd}$, and such a newly introduced phase can be used in explaining the large direct $CP$ asymmetries in $B \to K\pi$ \cite{Barger:2004hn,changkllzp}. Compared with the effective Hamiltonian of SM \cite{Buchalla}, the operators of the forms $({\bar b}d)_{V-A} ({\bar q}q)_{V-A}$ and $({\bar b}d)_{V-A} ({\bar q}q)_{V+A}$ in eq.(\ref{eqn:Heff1}) have already existed in SM, so the $Z^\prime$ effect can be represented by modifying the Wilson coefficients of the corresponding operators. Thus, the eq.(\ref{eqn:Heff1}) can be rewritten as
\begin{multline}
{\cal H}_{\rm eff}^{Z^\prime}=-\frac{G_F}{\sqrt{2}} V_{tb}V_{ts}^*\sum_q \left( \Delta C_3  O_3^{(q)} +\Delta C_5  O_5^{(q)} \right.\\
\left.+ \Delta C_7  O_7^{(q)} + \Delta C_9  O_9^{(q)} \right) +\mbox{h.c.},  \label{eqn:Heff2}
\end{multline}
where $O_{3,5,7,9}^{(q)}$ are the four quark operators  existing in SM\cite{Buchalla}. The additional contributions to the SM Wilson coefficients at the $M_W$ scale in terms of $Z^\prime$ parameters are thus given as
\begin{eqnarray}
& \Delta C _{3(5)}=- \frac{2}{3 V_{tb}V_{td}^* }\left(\frac{g_2 m_Z}{g_1m_{Z^\prime}}\right)^2 B^{L}_{bd} \left(B^{L(R)}_{uu} + 2 B^{L(R)}_{ss}\right),\\
& \Delta C _{9(7)} = -\frac{4}{3 V_{tb} V_{td}^*} \left(\frac{g_2m_Z}{g_1 m_{Z^\prime}}\right)^2 B^{L}_{bd} \left(B^{L(R)}_{uu} - B^{L(R)}_{ss} \right).
\end{eqnarray}
From above equations, we find that both the electro-weak penguins $\Delta C_{9(7)}$ and the QCD penguins $\Delta C_{3(5)}$ will be affected  by the new gauge boson $Z^\prime$. Since the scale of new physics is expected to be much higher than that of  electroweak scale, in order to show that the new physics is primarily manifest in the electroweak penguins, we therefore follow ref.\cite{Barger:2004hn,changkllzp,liy} to assume $B^{L(R)}_{uu} \simeq -2 B^{L(R)}_{ss}$. In this case, the $Z^\prime$ contributions to the Wilson coefficients are
\begin{eqnarray}
\Delta C_{3(5)}=0,\,\,\,
\Delta C_{9(7)}=4 \frac{|V_{tb}V_{td}^*|}{V_{tb}V_{td}^*}
\zeta^{L(R)} e^{i\phi_{bd}} ~,
\end{eqnarray}
where
\begin{eqnarray}\label{eqn:xi}
&&\zeta^{X}=\left(\frac{g_2 m_Z}{g_1 m_{Z^\prime}}\right)^2
\left|\frac{B^{L}_{bd} B^X_{ss}}{V_{tb}V_{td}^* }\right| ~~
(X=L,R),\\
&&\phi_{bd} ={\rm Arg}[B^L_{bd}] ~.
\end{eqnarray}
Finally, we obtain the resulting effective Hamiltonian at the $M_W$ scale
\begin{eqnarray}
{\cal H}_{\rm eff}^{Z^\prime}
&=& - \frac{4 G_F}{\sqrt{2}} \left(\frac{g_2 m_Z}{g_1 m_{Z^\prime}}\right)^2 B^{L*}_{bd}\sum_q \left( B^L_{ss} O_9^{(q)} + B^R_{ss} O_{7}^{(q)} \right) \nonumber \\
&&+ \mbox{h.c.}
\end{eqnarray}
Because all heavy degrees of freedom (including the $Z^\prime$) above the scale of  $W$ boson mass have been integrated out already and there is no new particles below $m_W$, the renormalization group evolution of above new Wilson coefficients down to low energies is exactly the same as in the SM \cite{Buchalla}.

Now, we will discuss the the ranges of new parameters $\zeta^{L,R}$ and $\phi_{bd}$. Because both the gauge group $U(1)_Y$  and  the new $U(1)^\prime$  are expected to origininate from the same large group, the relation $g_2/g_1 \sim 1$ is assumed. In the experimental side, if the leptophobic $Z^\prime$ boson is   detected at LHC, the mass of $Z^\prime$ should be about a few TeV, which means $ m_Z/m_{Z^\prime}$ is at the order of ${\cal O}(10^{-1})$. In addition, the other parameters $|B^{L}_{bd}|$, $|B^X_{qq}|$ and new weak phases $\phi_{bd}$ could be constrained by the data induced by FCNC. For instance, the mass difference $B_{d}^0$ and $\overline B_{d}^0$ requires $|B^{L}_{bd}|\sim|V_{tb}V_{td}^{*}|$. Then, with experimental data of $B_{d,s}$ nonleptonic charmless decays, $B^{L,R}_{qq}\sim 1$ could be extracted. As for the new introduced phase $\phi_{bd}$, recent discussions indicate that $\phi_{bd}=-50^\circ$ \cite{changkllzp}.   How to constrain these parameters globally is beyond the scope of this work. We will not discuss this topic explicitly here. So, to probe the new physics effect for maximum range, we assume  $\zeta = \zeta^{L}=\zeta^{R}\in [0.001,0.02]$, and $\phi_{bd} \in [-180^\circ, 180^\circ]$, and set $\zeta=0.01$ and $\phi_{bd}=-50^\circ$  for the center values.

\section{QCD factorization calculation}

In the $B \to  K_0^{*\pm}(1430) K^{(*)\mp}$  decays, none of the quarks in final states is the same as those of the initial $B$ meson. Therefore they can occur only via annihilation type diagrams.  However, in QCDF, the end-point singularity usually appears when calculating the annihilation type diagrams \cite{QCDF}. As a most popular way, the end-point divergent integral is treated as   infrared sensitive contribution that is parameterized by
\begin{equation}\label{treat-for-anni}
\int_0^1 \frac{\!dx}{1-x}\, \to X_A=\ln\left({m_B\over \Lambda}\right)(1+\rho_A e^{i\phi_A}),
\end{equation}
where the unknown parameters $\rho_A$ and $\phi_A$ could be fixed by the experimental data. This singularity can also be smeared by introducing the effective mass to gluon \cite{2006wc}. We will  adopt eq.(\ref{treat-for-anni}) in this work.

Within the framework of QCDF, the decay amplitudes of $B \to  K_0^{*\pm}(1430) K^{(*)\mp}$ can be written as
\end{multicols}
\ruleup
\begin{eqnarray} \label{eq:af5}
A(\overline B^0 \to  K_0^{*+} K^- ) &=&
 \frac{G_F}{\sqrt{2}}\sum_{p=u,c}\lambda_p^{(d)}
 \Bigg\{  f_Bf_{K_0^*}f_K\big[\left(b_1\delta_u^p+b_4
 +b_{\rm 4,EW}\right)_{K_0^{*+} K^-}+\left(b_4
 -\frac{1}{2}b_{\rm 4,EW}\right)_{K^- K_0^{*+} }\big] \Bigg\},
\end{eqnarray}
\begin{eqnarray} \label{eq:af6}
A(\overline B^0 \to K^+ K_0^{*-}) &=&
 \frac{G_F}{\sqrt{2}}\sum_{p=u,c}\lambda_p^{(d)}
 \Bigg\{
  f_Bf_{K_0^*}f_K\big[\left(b_1\delta_u^p+b_4
 +b_{\rm 4,EW}\right)_{K^+ K_0^{*-}}+\left(b_4
 -\frac{1}{2}b_{\rm 4,EW}\right)_{K_0^{*-}K^+ }\big] \Bigg\},
\end{eqnarray}
\begin{eqnarray} \label{eq:af11}
A(\overline B^0 \to  K_0^{*+} K^{*-} ) &=&
 \frac{G_F}{\sqrt{2}}\sum_{p=u,c}\lambda_p^{(d)}
 \Bigg\{
  -f_Bf_{K_0^*}f_{K^*}\big[\left(b_1\delta_u^p+b_4
 +b_{\rm 4,EW}\right)_{K_0^{*+} K^{*-}}+\left(b_4
 -\frac{1}{2}b_{\rm 4,EW}\right)_{K^{*-} K_0^{*+} }\big] \Bigg\},
\end{eqnarray}
\begin{eqnarray} \label{eq:af12}
A(\overline B^0 \to K^{*+} K_0^{*-}) &=&
 \frac{G_F}{\sqrt{2}}\sum_{p=u,c}\lambda_p^{(d)}
 \Bigg\{-f_Bf_{K_0^*}f_{K^*}\big[\left(b_1\delta_u^p+b_4
 +b_{\rm 4,EW}\right)_{K^{*+} K_0^{*-}}+\left(b_4
 -\frac{1}{2}b_{\rm 4,EW}\right)_{K_0^{*-}K^{*+} }\big] \Bigg\},
\end{eqnarray}
\ruledown
\begin{multicols}{2}
where the building blocks $b_i$ and $b_{i,EW}$ read
\begin{eqnarray}
 b_1 &=& {C_F\over N_c^2}C_1A_1^i, \nonumber\\
 b_2 &=& {C_F\over N_c^2}C_2A_1^i, \nonumber \\
 b_3 &=&{C_F\over
 N_c^2}\left[C_3^\prime A_1^i+C_5^\prime (A_3^i+A_3^f)+N_cC_6^\prime A_3^f\right], \nonumber \\
 b_4&=&{C_F\over
 N_c^2}\left[C_4^\prime A_1^i+C_6^\prime A_2^f\right], \nonumber \\
 b_{\rm 3,EW} &=& {C_F\over
 N_c^2}\left[C_9^\prime A_1^{i}+C_7^\prime (A_3^{i}+A_3^{f})+N_cC_8^\prime A_3^{i}\right],
\nonumber \\
 b_{\rm 4,EW} &=& {C_F\over
 N_c^2}\left[C_{10}^\prime A_1^{i}+C_8^\prime A_2^{i}\right].
\end{eqnarray}
The expressions of the functions $A_n^{i,f}$ can be found in Ref.\cite{Cheng:2005nb,Cheng:2007st}, and the subscripts 1,2,3 denote the annihilation amplitudes induced from $(V-A)(V-A)$, $(V-A)(V+A)$ and $(S-P)(S+P)$ operators, while the superscripts $i$ and $f$ refer to gluon emission from the initial and final-state quarks, respectively. When listing the two final state mesons  $M_1M_2$ in the formulas, we refer the first meson $M_1$ to containing an anti-quark from the weak vertex and refer $M_2$ to containing a quark from the weak vertex. As for the aforementioned endpoint singularity $X_A$, we  adopt the eq.(\ref{treat-for-anni}) with $\Lambda =0.5 \mathrm{GeV}$. Note that the recent global fit of $\rho_A$ and $\phi_A$ to $B\to SP, SV$ implies $\rho_A=0.15$ and $\phi_A = 82^\circ$ with $\chi^2= 8.3$ \cite{Cheng:2013fba}. Therefore, we set $\rho_A\in [0.1, 0.2] $ and $\phi_A \in [60^\circ, 120^\circ]$ in estimating the uncertainties.

Because both $B^0$ and $\overline B^0$ could decay to $K_0^{*+}(1430)K^{-}$ and $K_0^{*-}(1430) K^{+}$,  we then define four decay amplitudes, $A_f$, $A_{\bar f}$, $\bar A_f$ and $\bar A_{\bar f}$ as
\begin{eqnarray}
& A_{f}=\langle K_0^{*+}  K^-|B^0\rangle,
 A_{\bar f}=\langle K_0^{*-}  K^+|B^0\rangle;\nonumber \\
& \bar A_{f}=\langle K_0^{*+}  K^-|\overline B^0\rangle,
 \bar A_{\bar f}=\langle K_0^{*-}  K^+|\overline B^0\rangle.
\end{eqnarray}
Then, the direct $CP$ asymmetry is defined as
\begin{eqnarray}
 A_{CP}=\frac{|A_f|^2+|\bar A_f|^2-|A_{\bar f}|^2-|\bar A_{\bar f}|^2}{|A_f|^2+|\bar A_f|^2+|A_{\bar f}|^2+|\bar A_{\bar f}|^2}.
\end{eqnarray}
\end{multicols}

\begin{table*}[h]
\begin{center}
\caption{Branching fractions of $B \to K_0^*(1430)K^{(*)}$ under different scenarios.}
\label{Table:br}
\begin{tabular}{c|c|c|c}
\hline
\hline
Scenario
&Decay Modes
&SM
&$Z^\prime$
 \\\hline
\multirow{4}{2cm}{Scenario-1}
&$\overline B^0\to K_0^{*+}(1430)K^-$
&$0.97^{+0.43+0.23}_{-0.31-0.12}$
&$1.37^{+0.57+0.35+0.62}_{-0.43-0.19-1.05}$
\\\cline{2-4}
&$\overline B^0\to K_0^{*-}(1430)K^+$
&$8.33^{+3.31+2.65}_{-2.55-1.39}$
&$8.17^{+3.25+2.60+0.59}_{-2.50-1.36-0.18}$\\
\cline{2-4}
&$\overline B^0\to K_0^{*+}(1430)K^{*-}$
&$0.33^{+0.14+0.08}_{-0.10-0.04}$
&$0.40^{+0.16+0.10+0.10}_{-0.12-0.05-0.20}$
\\\cline{2-4}
&$\overline B^0\to K_0^{*-}(1430)K^{*+}$
&$14.54^{+5.75+4.54}_{-4.44-2.38}$
&$14.47^{+5.72+4.52+2.67}_{-4.42-2.37-0.08}$\\
\hline
\multirow{4}{2cm}{Scenario-2}
&$\overline B^0\to K_0^{*+}(1430)K^-$
&$0.58^{+0.45+0.04}_{-0.29-0.02}$
&$0.65^{+0.50+0.08+0.07}_{-0.32-0.03-0.18}$
\\\cline{2-4}
&$\overline B^0\to K_0^{*-}(1430)K^+$
&$1.07^{+0.72+0.79}_{-0.47-0.29}$
&$1.06^{+0.71+0.77+0.08}_{-0.47-0.28-0.06}$\\
\cline{2-4}
&$\overline B^0\to K_0^{*+}(1430)K^{*-}$
&$0.11^{+0.09+0.03}_{-0.06-0.02}$
&$0.12^{+0.09+0.04+0.01}_{-0.06-0.02-0.02}$
\\\cline{2-4}
&$\overline B^0\to K_0^{*-}(1430)K^{*+}$
&$1.84^{+1.26+1.38}_{-0.83-0.52}$
&$1.84^{+1.25+1.37+0.04}_{-0.82-0.51-0.04}$\\
\hline
\hline
\end{tabular}
\end{center}
\end{table*}

\begin{multicols}{2}
In the experimental side, the four time-dependent decay widths are given by ($f=K_0^{*+}  K^-$ and $ \bar f=K_0^{*-}  K^+$)
\begin{eqnarray}\label{eq:shuai}
\Gamma(B^0(t)\to f)&=&e^{-\Gamma t}\frac{1}{2}(|A_f|^2+|\bar{A}_f|^2)\nonumber\\
&&[1+C_f\cos\Delta m t-S_f\sin\Delta m t],
\nonumber\\
\Gamma(\overline{B}^0(t)\to \bar f)&=&e^{-\Gamma t}\frac{1}{2}(|A_{\bar{f}}|^2+|\bar{A}_{\bar{f}}|^2)
\nonumber\\
&&[1-C_{\overline{f}}\cos\Delta m t+S_{\bar{f}}\sin\Delta m t],\nonumber\\
\Gamma(B^0(t)\to \bar f)&=&e^{-\Gamma t}\frac{1}{2}(|A_{\bar{f}}|^2+|\bar{A}_{\bar{f}}|^2)
\nonumber\\
&&[1+C_{\overline{f}}\cos\Delta m t-S_{\bar{f}}\sin\Delta m t],\nonumber\\
\Gamma(\overline{B}^0(t)\to  f)&=&e^{-\Gamma t}\frac{1}{2}(|A_f|^2+|\bar{A}_f|^2)\nonumber\\
&&[1-C_f\cos\Delta m t+S_f
\sin\Delta m t].
\end{eqnarray}
In above functions, $\Delta m$ means the mass difference of $B^0/\overline B^0$ meson, and $\Gamma$ is the average decay width of the $B$ meson. The auxiliary parameters $C_f$ and $S_f$, that can be extracted from the data, are given by
\begin{eqnarray}
C_f&=&\frac{|A_f|^2-|\bar{A}_f|^2}{|\bar{A}_f|^2+|A_f|^2}, \\ S_f&=&\frac{2\mathrm{Im}(\lambda_f)}{1+|\bar{A}_f/A_f|^2},\\
\lambda_f&=&\frac{V_{tb}V_{td}^*}{V_{tb}^*V_{td}}\frac{\bar{A}_f}{A_f}.
\end{eqnarray}
The definitions of $C_{\bar f}$ and $S_{\bar f}$ can also been obtained by replacing $f$ with $\bar{f}$. In order to show the implications of above four parameters, we usually use the following four new parameters:
\begin{eqnarray}
C=\frac{1}{2}(C_f+C_{\bar{f}}), &&\Delta C=\frac{1}{2}(C_f-C_{\bar{f}}),\\
S=\frac{1}{2}(S_f+S_{\bar{f}}), &&\Delta S=\frac{1}{2}(S_f-S_{\bar{f}}).
\end{eqnarray}
Physically, $S$ and $C$ are related to the mixing-induced $CP$ asymmetry and the direct $CP$ asymmetry, respectively. Moreover, $\Delta C$ and $\Delta S$ are $CP$-even under $CP$ transformation $\lambda_f\rightarrow 1/\lambda_{\bar{f}}$.

\section{Numerical Results}

Using the parameters of refs.\cite{Cheng:2005nb, Cheng:2007st,Cheng:2013fba}, with and without $Z^\prime$ boson, we present the predicted branching ratios of $B \to  K_0^{*\pm}(1430)K^{(*)\mp}$ under different scenarios ($K_0^{*\pm}(1430)$ as the first excited states   or the ground states) in Table.\ref{Table:br}. The predictions of SM are also listed for comparison.
Within the PQCD approach based on the $k_T$ factorization, Liu Xin $et.al.$ had studied these decay modes \cite{Liuxin}. Comparing their results with ours, we find that their results of branching ratios are much larger than ours by $1\sim 2$ orders of magnitude. Currently, we cannot determine which approach is better, but we hope the future experiment can test these two different approaches.

From Table.\ref{Table:br}, for decay modes $\overline B^0\to K_0^{*+}(1430)K^{(*)-}$, the differences between S1 and S2 is very small, so it is hard for us to discriminate two different scenarios using these two decays. In contrast, the branching fractions of  $\overline B^0\to K_0^{*-}(1430)K^{(*)+}$ under S1 and S2 have large differences, so if these two modes can be measured precisely they may be used to determine whether $K_0(1430)$ belongs to the ground states or the first-excited states.

In Table.\ref{Table:cp}, we also list our predictions of $A_{CP}$, $C$, $\Delta C$, $S$ and $\Delta S$ for the final states $ K_0^{*+}K^-$ and $K_0^{*+}K^{*-}$, under two different scenarios in both SM (the upper values) and $Z^\prime$ model (the lower values). For all results, the first errors are from the uncertainties of decay constants and the light-cone distribution amplitudes of final states and the second errors come from the $\rho_A$ and $\phi_A$, and the last errors in the $Z^\prime$ model are the results by scanning the possible ranges of $\zeta$ and $\phi_{bd}$.  From the numerical results, we find that the largest uncertainties are from the $\rho_A$ and $\phi_A$. This is in contrast with other decay modes dominated  by the spectator diagrams, whose   uncertainties taken by the two above parameters are small.  Unlike branching fractions, the  $CP$ asymmetry parameters are not sensitive to non-perturbative  hadronic parameters, where these uncertainties   are canceled because they are ratios of different amplitudes. Therefore, these parameters are more sensitive to   the effect of NP.

\end{multicols}

\begin{table*}[htb!]
\begin{center}
\caption{The CP-violating parameters $A_{CP}$, $C$, $\Delta C$, $S$  and $\Delta S$  ($\%$) of $B \to K_0^*(1430)K^{(*)}$.}
\label{Table:cp}
\begin{tabular}{c|c|l|l|l|l|l}
\hline
\hline
Decay
&S.
&$A_{CP}$
&$C$
&$\Delta C$
&$S$
&$\Delta S$ \\\hline
\multirow{4}{1.2cm}{$ K_0^{*\pm}K^{\mp}$}&
\multirow{2}{0.5cm}{$S1$}
&$ 0.79^{+0.01+0.01}_{-0.01-0.01}$
&$-0.04^{+0.01+0.02}_{-0.01-0.01}$
&$ 0.06^{+0.01+0.02}_{-0.02-0.03}$
&$ 0.04^{+0.02+0.01}_{-0.02-0.01}$
&$-0.84^{+0.03+0.02}_{-0.03-0.02}$\\&
&$ 0.71^{+0.01+0.01+0.22}_{-0.01-0.01-0.11}$
&$-0.02^{+0.00+0.00+0.05}_{-0.00-0.01-0.11}$
&$ 0.01^{+0.01+0.02+0.22}_{-0.01-0.03-0.12}$
&$ 0.04^{+0.02+0.01+0.02}_{-0.02-0.01-0.07}$
&$-0.88^{+0.02+0.01+0.19}_{-0.02-0.02-0.03}$\\
\cline{2-7}
&
\multirow{2}{0.5cm}{$S2$}
&$ 0.29^{+0.06+0.20}_{-0.09-0.14}$
&$-0.02^{+0.04+0.04}_{-0.02-0.05}$
&$ 0.25^{+0.04+0.09}_{-0.02-0.12}$
&$-0.36^{+0.05+0.06}_{-0.05-0.04}$
&$-0.20^{+0.12+0.10}_{-0.10-0.16}$\\&
&$ 0.24^{+0.07+0.18+0.14}_{-0.07-0.12-0.05}$
&$ 0.02^{+0.03+0.04+0.08}_{-0.01-0.05-0.19}$
&$ 0.27^{+0.06+0.12+0.10}_{-0.05-0.15-0.16}$
&$-0.35^{+0.06+0.06+0.08}_{-0.06-0.04-0.21}$
&$-0.26^{+0.14+0.11+0.19}_{-0.11-0.18-0.05}$\\
\hline
\multirow{4}{1.2cm}{$ K_0^{*\pm}K^{*\mp}$}&
\multirow{2}{0.5cm}{$S1$}
&$ 0.96^{+0.00+0.00}_{-0.00-0.00}$
&$-0.01^{+0.00+0.01}_{-0.00-0.00}$
&$ 0.02^{+0.00+0.01}_{-0.00-0.01}$
&$ 0.01^{+0.00+0.00}_{-0.00-0.00}$
&$-0.98^{+0.00+0.02}_{-0.00-0.00}$\\&
&$ 0.94^{+0.00+0.00+0.03}_{-0.00-0.00-0.01}$
&$-0.01^{+0.00+0.01+0.04}_{-0.00-0.00-0.02}$
&$ 0.01^{+0.00+0.01+0.07}_{-0.00-0.01-0.04}$
&$ 0.01^{+0.00+0.00+0.01}_{-0.00-0.00-0.00}$
&$-0.98^{+0.00+0.00+0.02}_{-0.00-0.00-0.01}$\\
\cline{2-7}
&
\multirow{2}{0.5cm}{$S2$}
&$ 0.88^{+0.02+0.03}_{-0.01-0.02}$
&$ 0.01^{+0.01+0.02}_{-0.01-0.02}$
&$ 0.11^{+0.03+0.06}_{-0.02-0.06}$
&$ 0.02^{+0.01+0.08}_{-0.01-0.04}$
&$-0.91^{+0.02+0.02}_{-0.02-0.02}$\\&
&$ 0.87^{+0.02+0.03+0.02}_{-0.01-0.02-0.01}$
&$ 0.03^{+0.02+0.03+0.02}_{-0.01-0.02-0.06}$
&$ 0.14^{+0.04+0.07+0.02}_{-0.03-0.07-0.08}$
&$ 0.02^{+0.01+0.01+0.02}_{-0.02-0.01-0.02}$
&$-0.92^{+0.03+0.03+0.02}_{-0.02-0.03-0.02}$\\
\hline
\hline
\end{tabular}
\end{center}
\end{table*}

\begin{multicols}{2}

\begin{figure}[t]
\begin{center}
\includegraphics[scale=0.7]{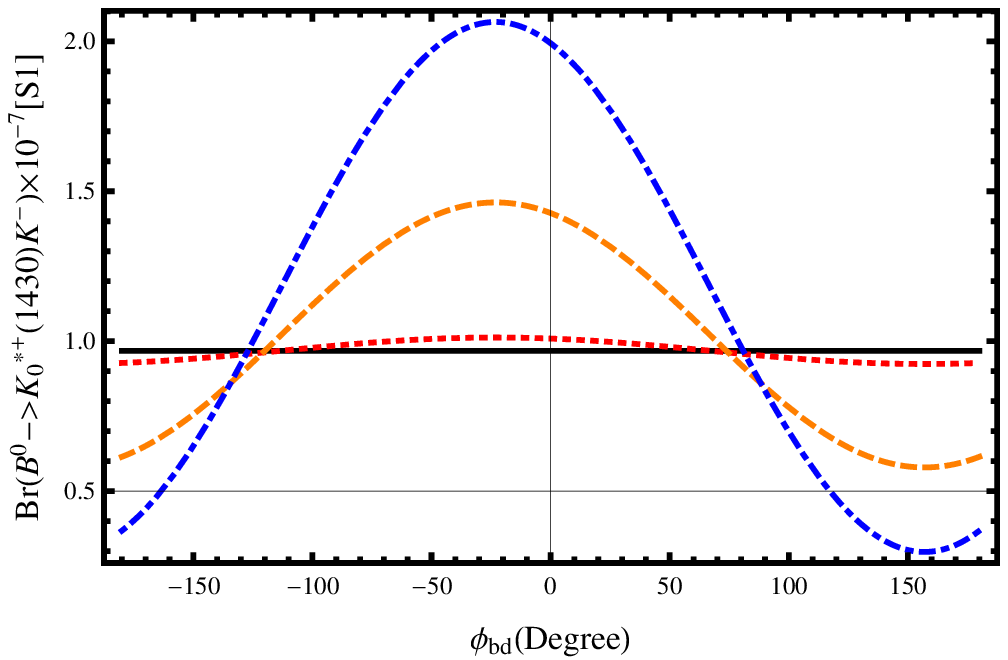}\hspace{2cm}
\includegraphics[scale=0.7]{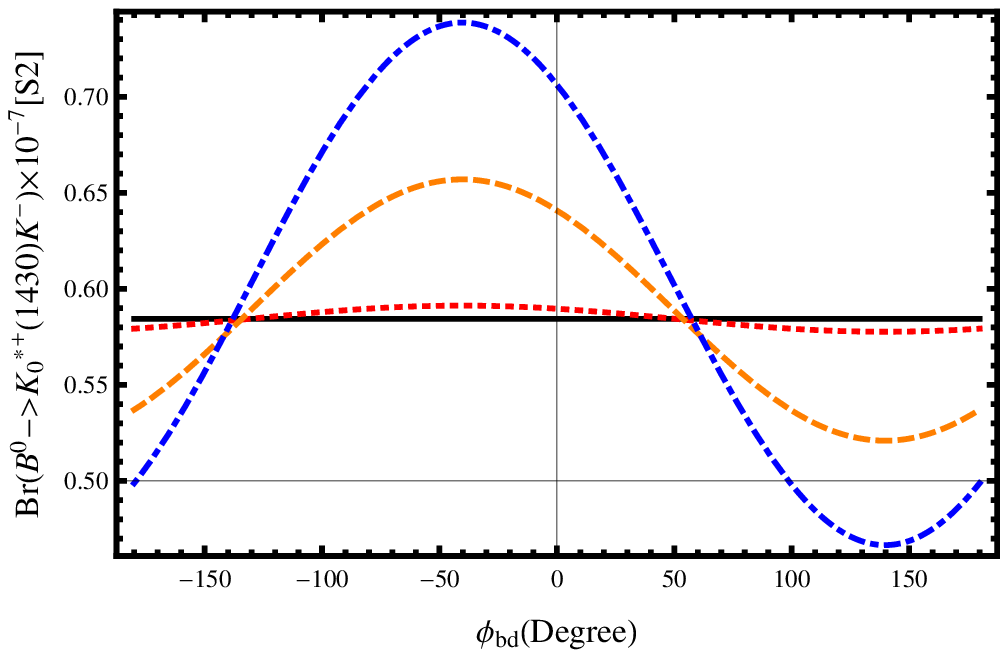}\\
\includegraphics[scale=0.7]{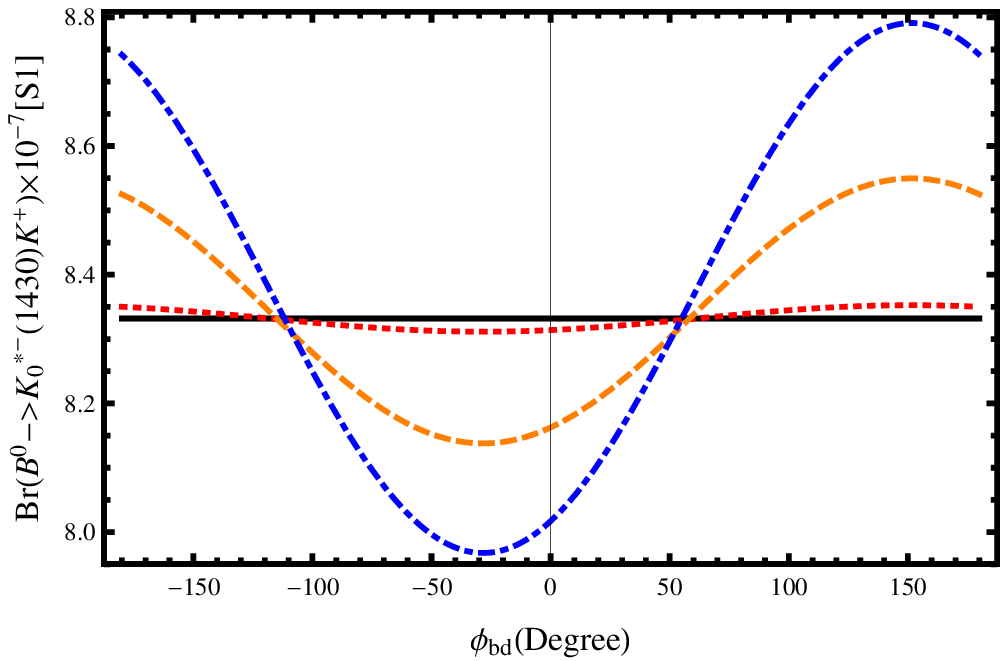}\hspace{2cm}
\includegraphics[scale=0.7]{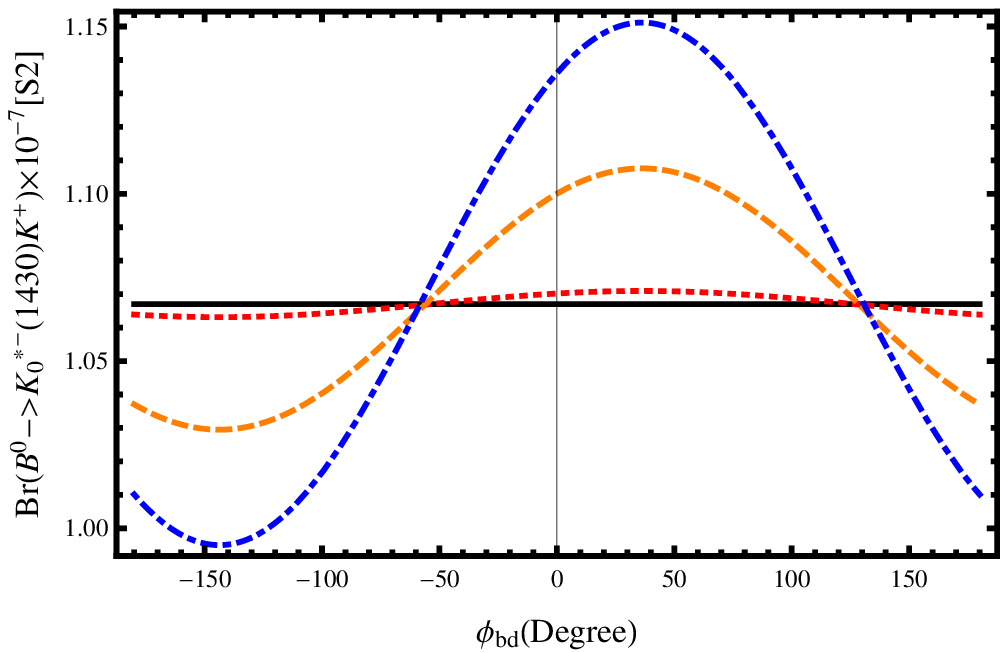}\\
\includegraphics[scale=0.7]{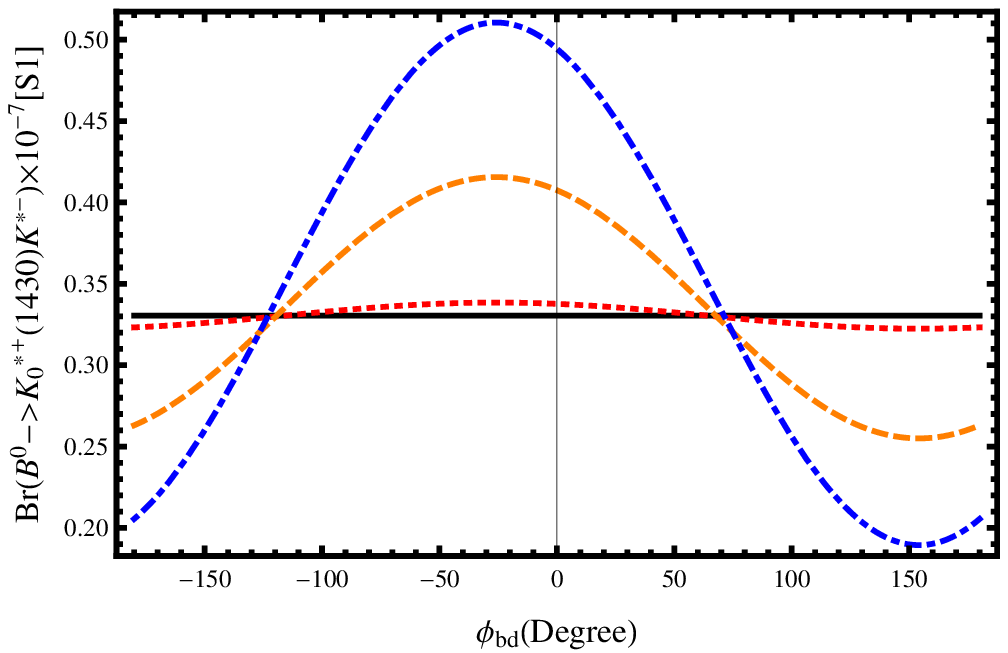}\hspace{2cm}
\includegraphics[scale=0.7]{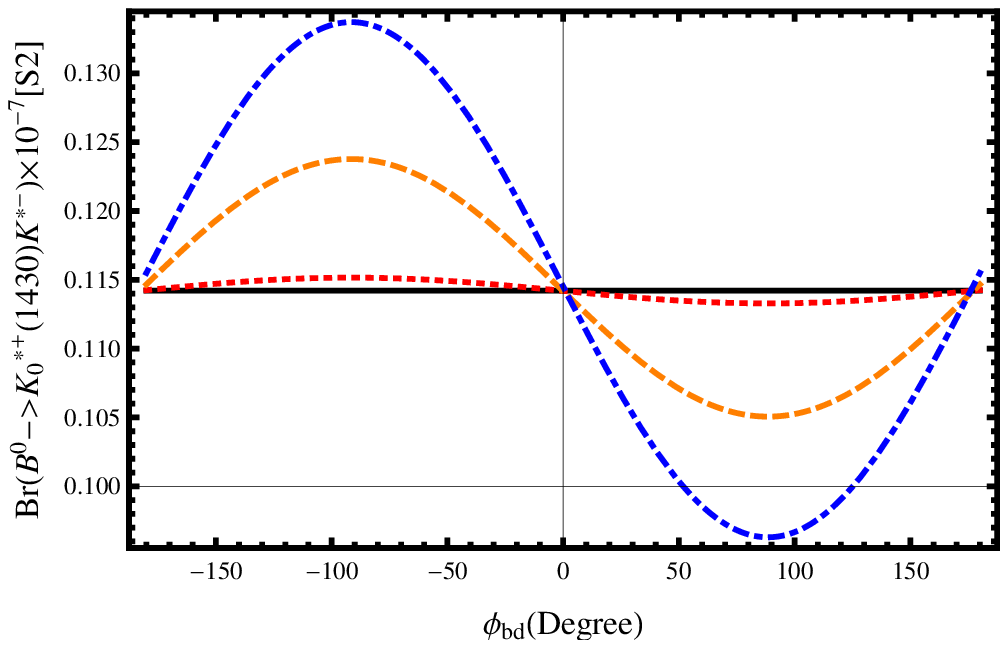}\\
\includegraphics[scale=0.7]{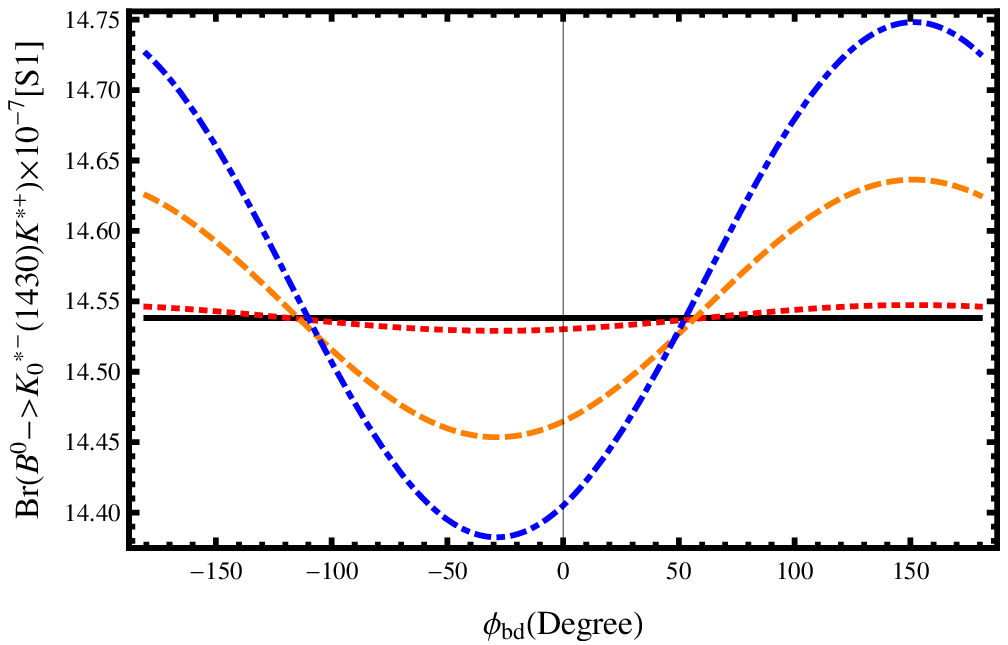}\hspace{2cm}
\includegraphics[scale=0.7]{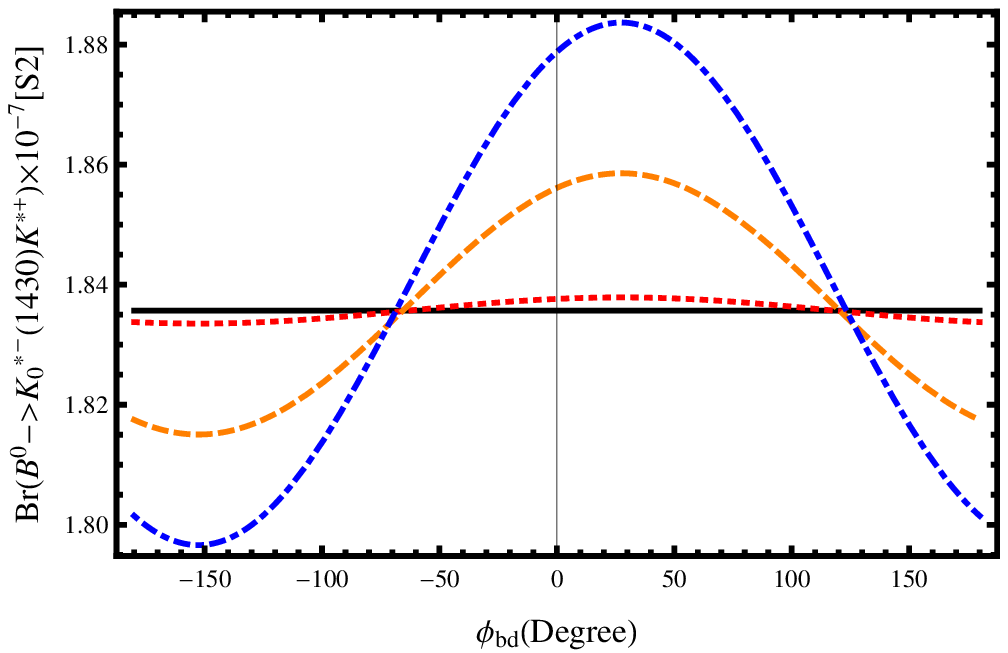}
\caption{Under different scenarios(S1 and S2), the branching ratios of $B \to K_0^{*\pm}(1430)K^{(*)\mp}$ as  functions of the weak phase $\phi_{bd}$, the dotted (red), dashed (orange) and dotdashed (blue) lines represent results from the $\zeta=0.001,~0.01,~0.02$, and the solid lines (black) are the predictions of SM.} \label{Fig:br}
 \end{center}
 \end{figure}

In order to show the effects of $Z^\prime$ clearly, we also plot the variations of the branching ratios as functions of the new weak phase $\phi_{bd}$ with different $\zeta=0.001,0.01,0.02$ under different scenarios in Fig.\ref{Fig:br}.
 Including the newly introduced $Z^\prime$ boson, one can see that if $\zeta <0.01$ the effects of $Z^\prime$ are not large enough to be detected in these four decay modes, because the new physics contributions are buried by the  large uncertainties of hadronic parameters.  If $\zeta$ is around 0.2, the branching fraction of $\overline B^0\to K_0^{*+}(1430)K^-$ under S1 will be changed remarkably to reach to $2.0\times 10^{-7}$, which could be measured in the forthcoming Belle-II experiment.

\begin{figure}[t]
\begin{center}
\includegraphics[scale=0.65]{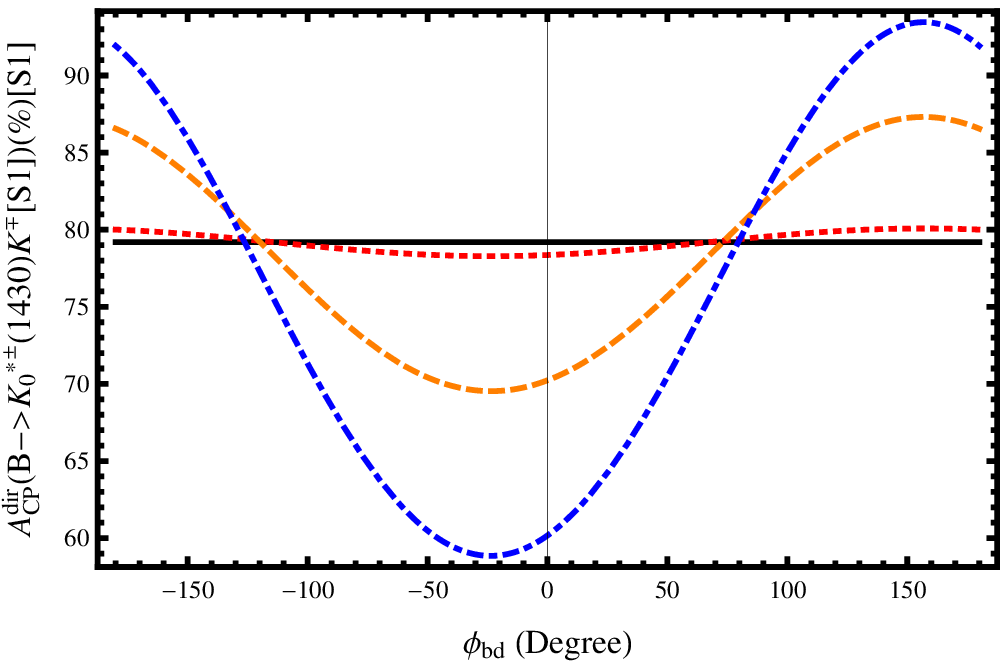}\hspace{2cm}
\includegraphics[scale=0.65]{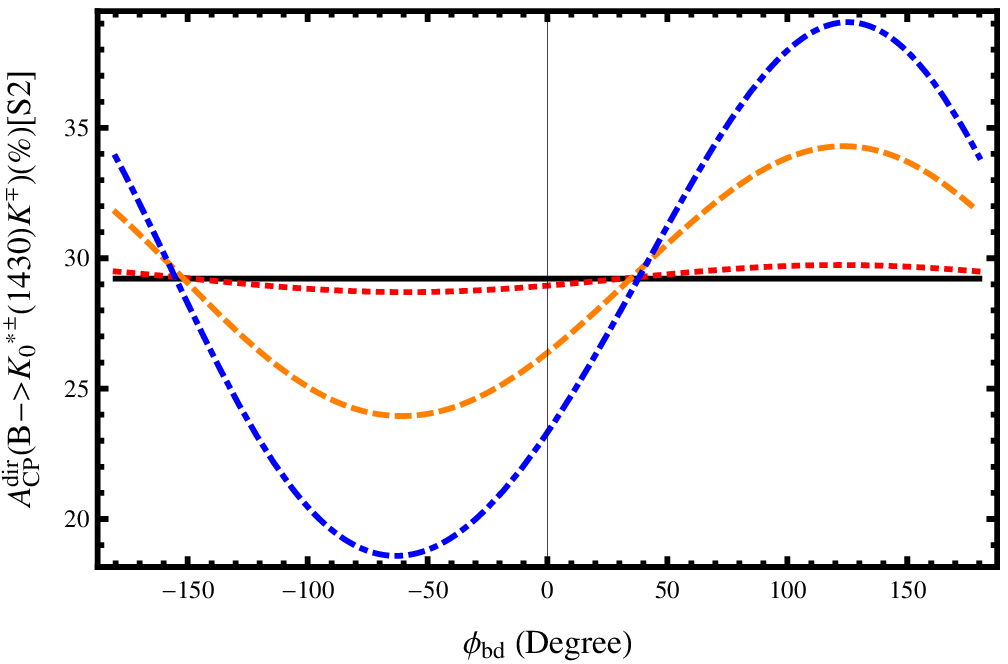}\\
\includegraphics[scale=0.65]{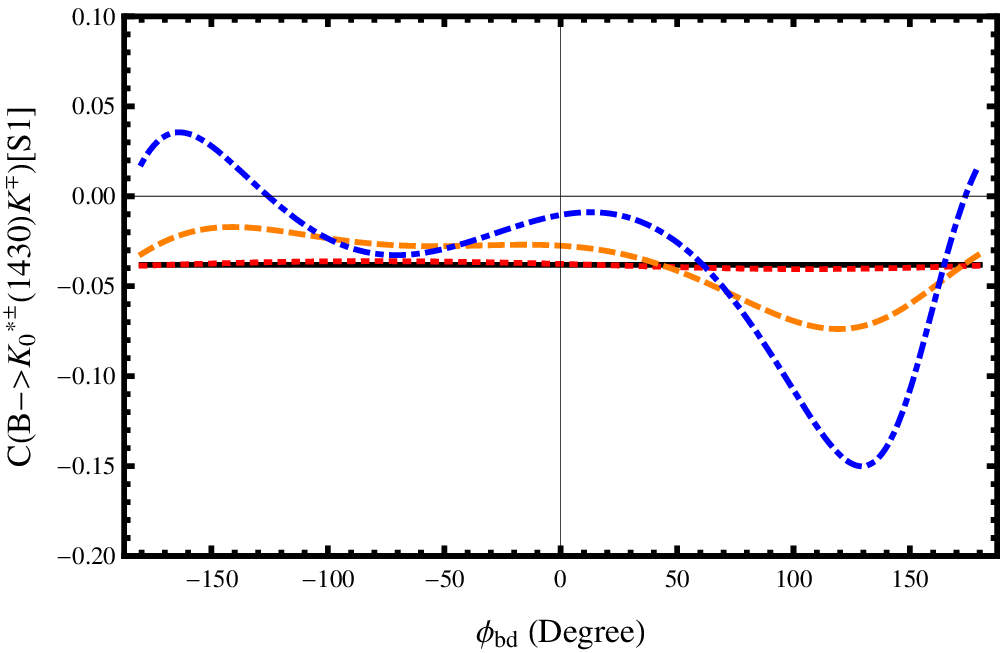}\hspace{2cm}
\includegraphics[scale=0.65]{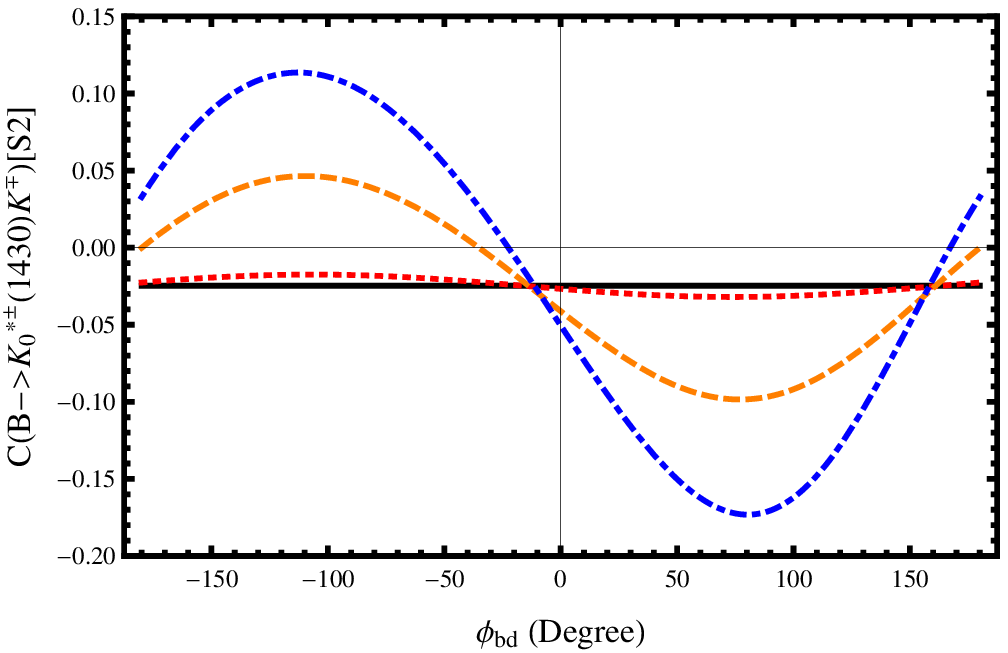}\\
\includegraphics[scale=0.65]{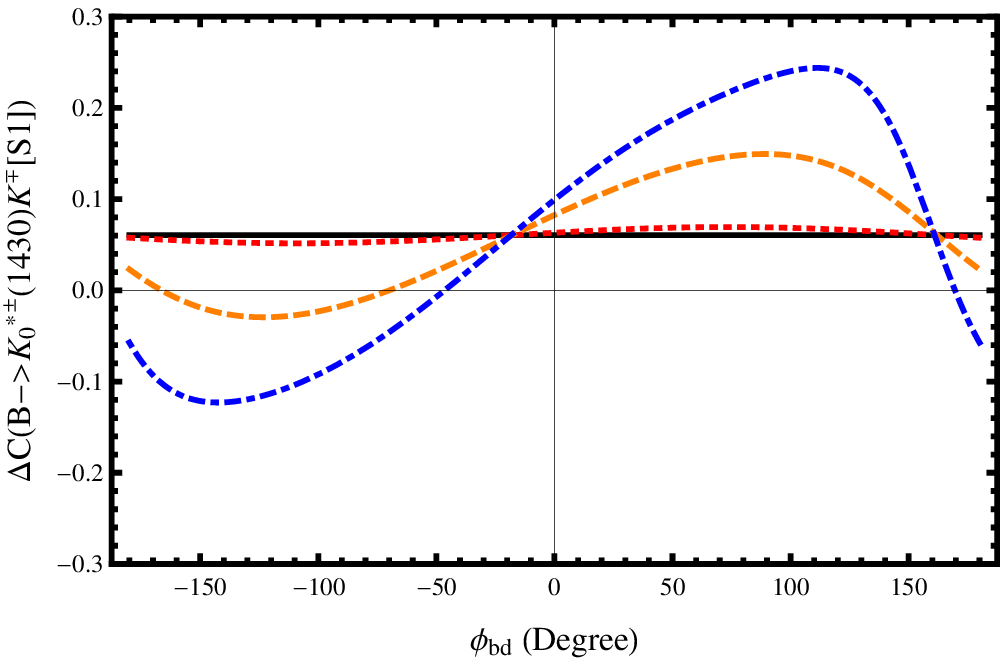}\hspace{2cm}
\includegraphics[scale=0.65]{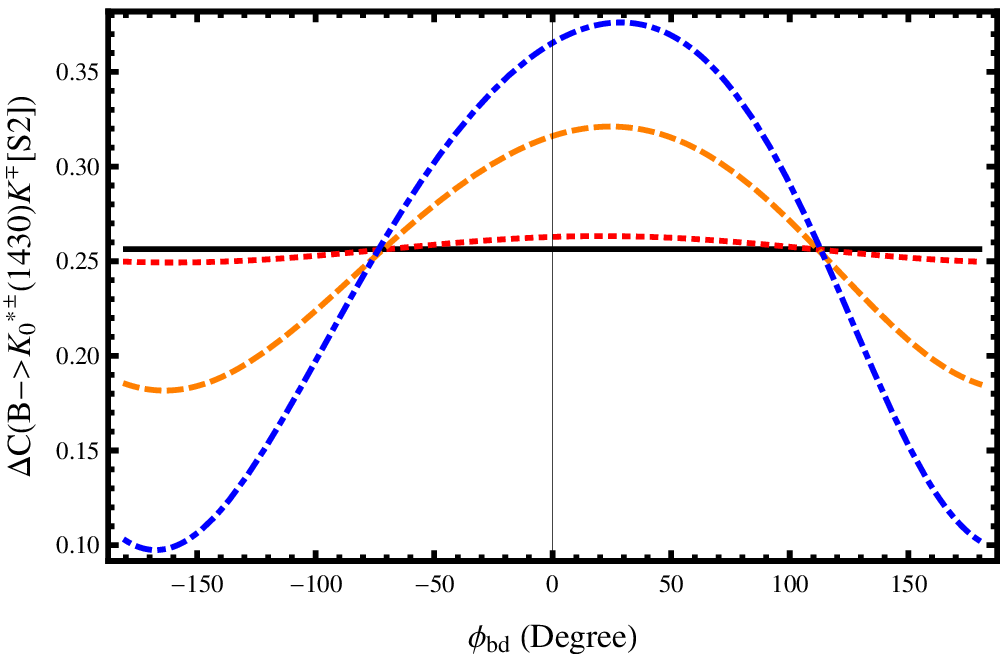}\\
\includegraphics[scale=0.65]{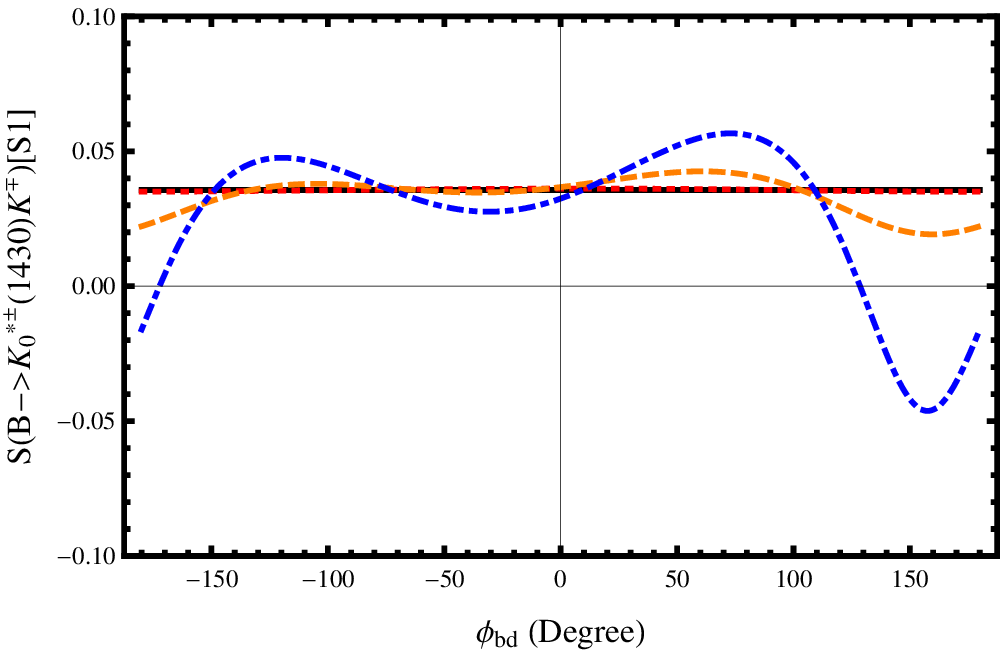}\hspace{2cm}
\includegraphics[scale=0.65]{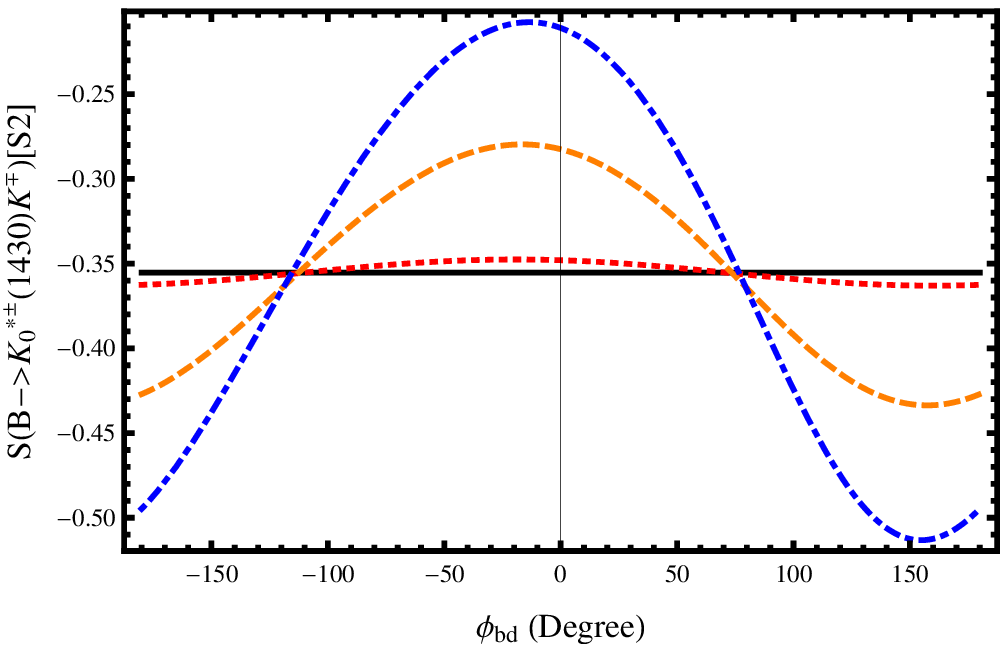}\\
\includegraphics[scale=0.65]{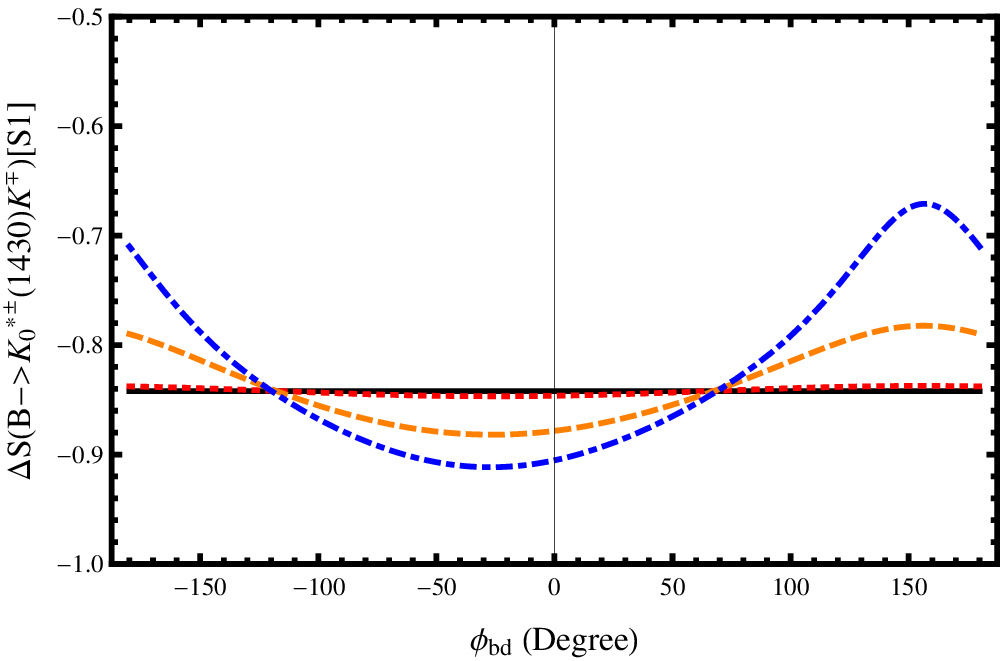}\hspace{2cm}
\includegraphics[scale=0.65]{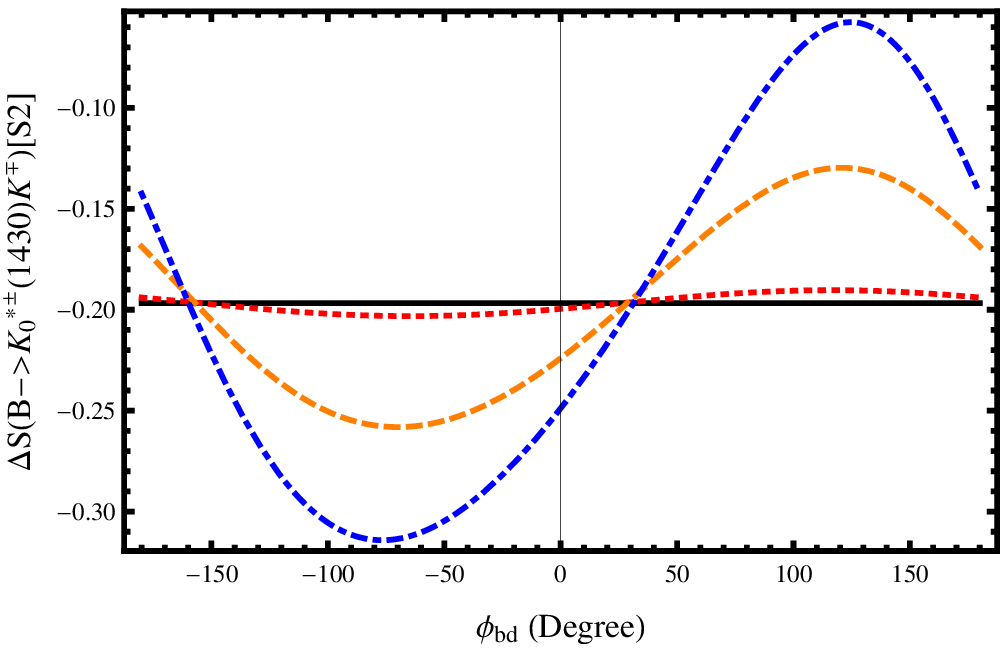}
\caption{Under different scenarios(S1 and S2), the CP-violating parameters $A_{CP}$, $C$, $\Delta C$, $S$  and $\Delta S$  ($\%$) of $B \to K_0^*(1430)K$ as functions of the weak phase $\phi_{bd}$, the dotted (red), dashed (orange) and dot-dashed (blue) lines represent results from the $\zeta=0.001,~0.01,~0.02$, and the solid lines (black) are the predictions of SM.} \label{Fig:cpsp}
 \end{center}
 \end{figure}

\begin{figure}[t]
\begin{center}
\includegraphics[scale=0.65]{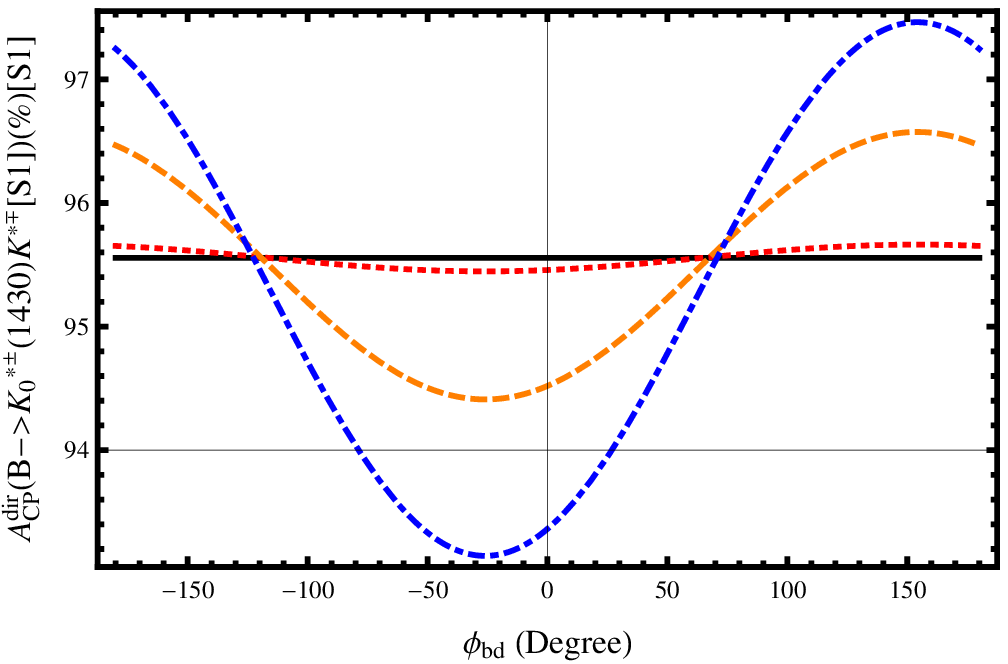}\hspace{2cm}
\includegraphics[scale=0.65]{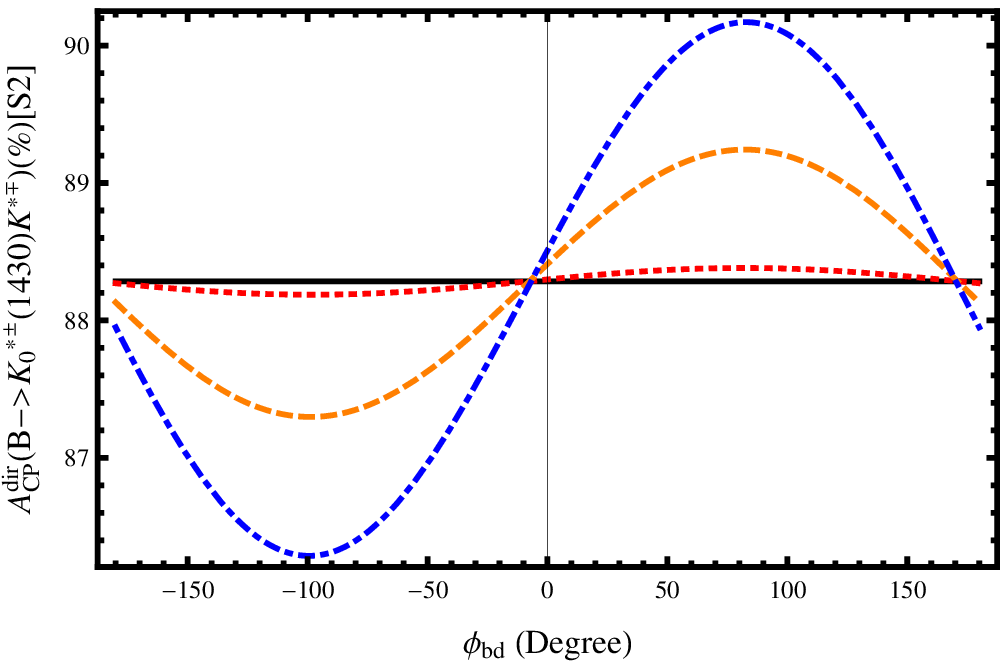}\\
\includegraphics[scale=0.65]{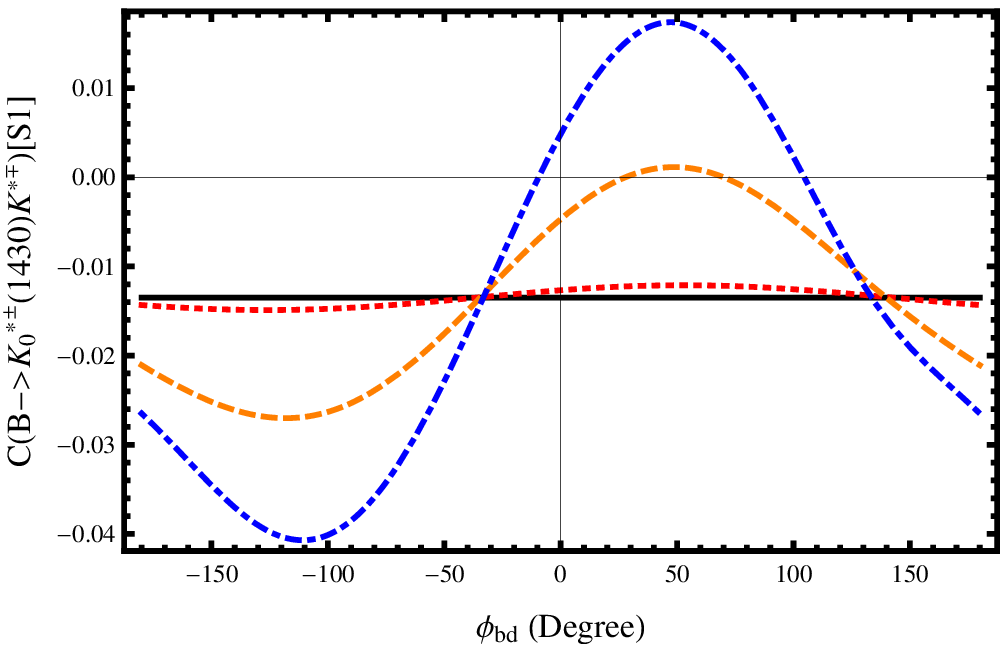}\hspace{2cm}
\includegraphics[scale=0.65]{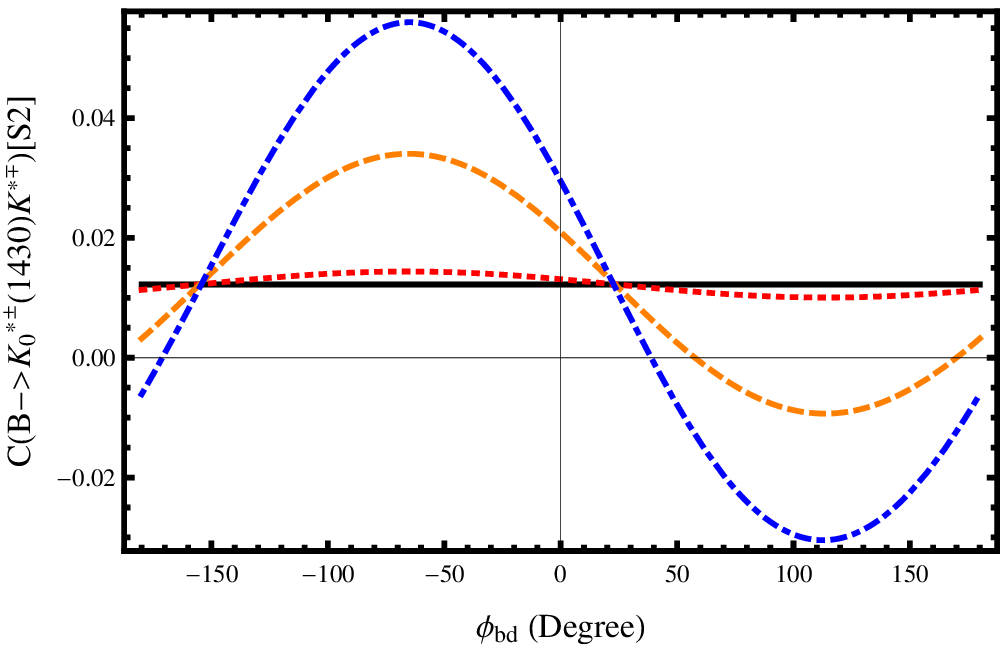}\\
\includegraphics[scale=0.65]{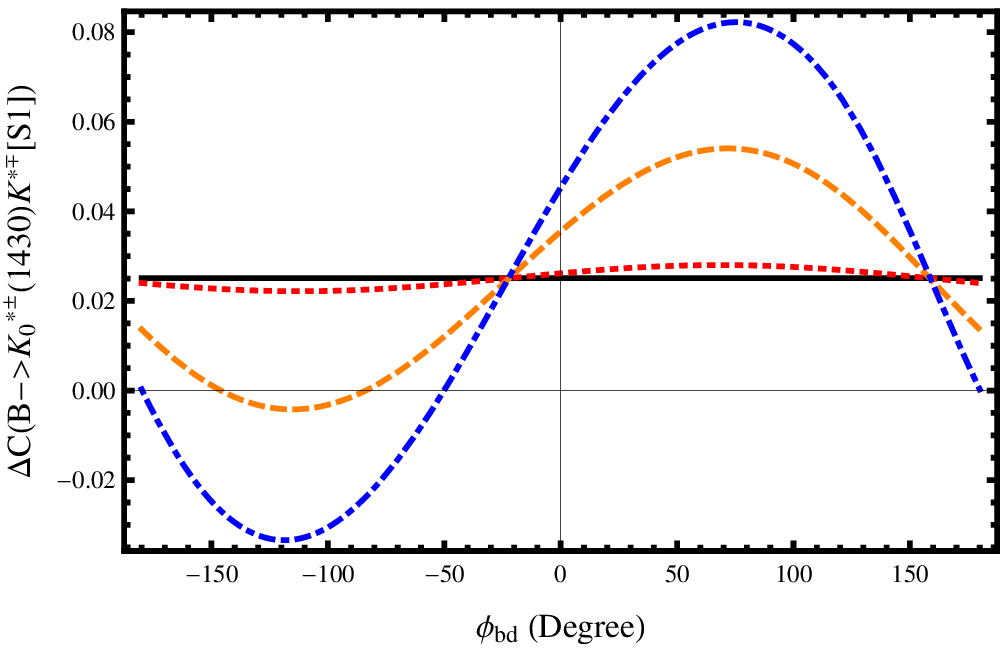}\hspace{2cm}
\includegraphics[scale=0.65]{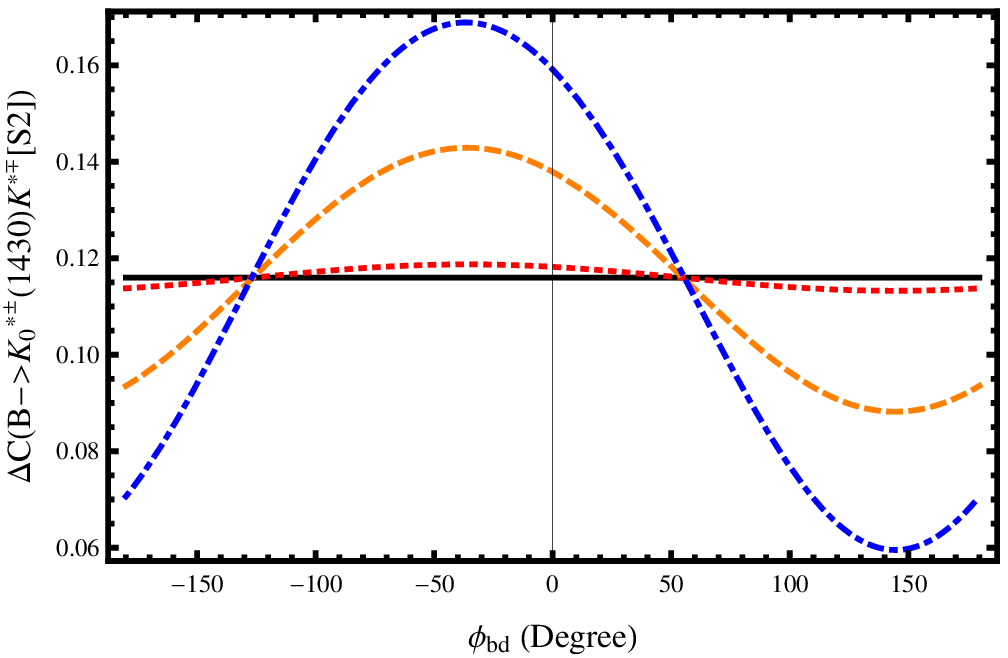}\\
\includegraphics[scale=0.65]{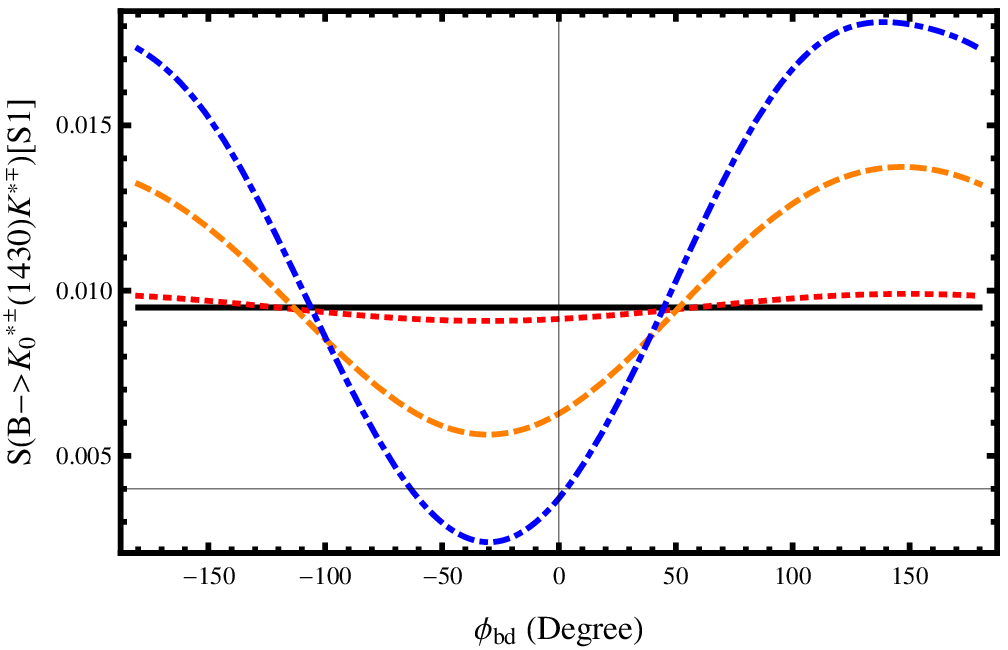}\hspace{2cm}
\includegraphics[scale=0.65]{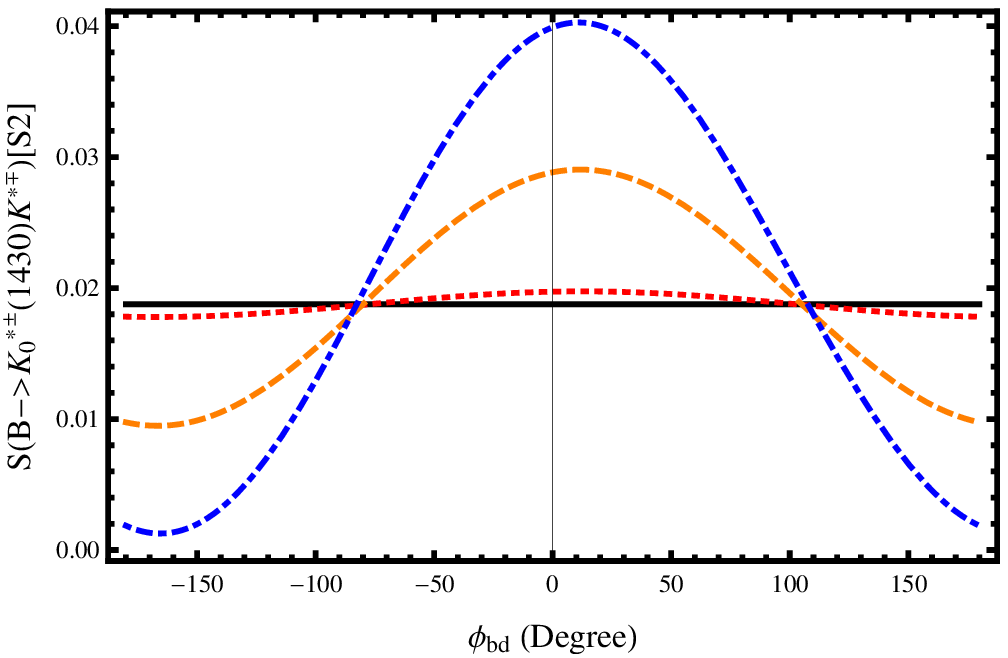}\\
\includegraphics[scale=0.65]{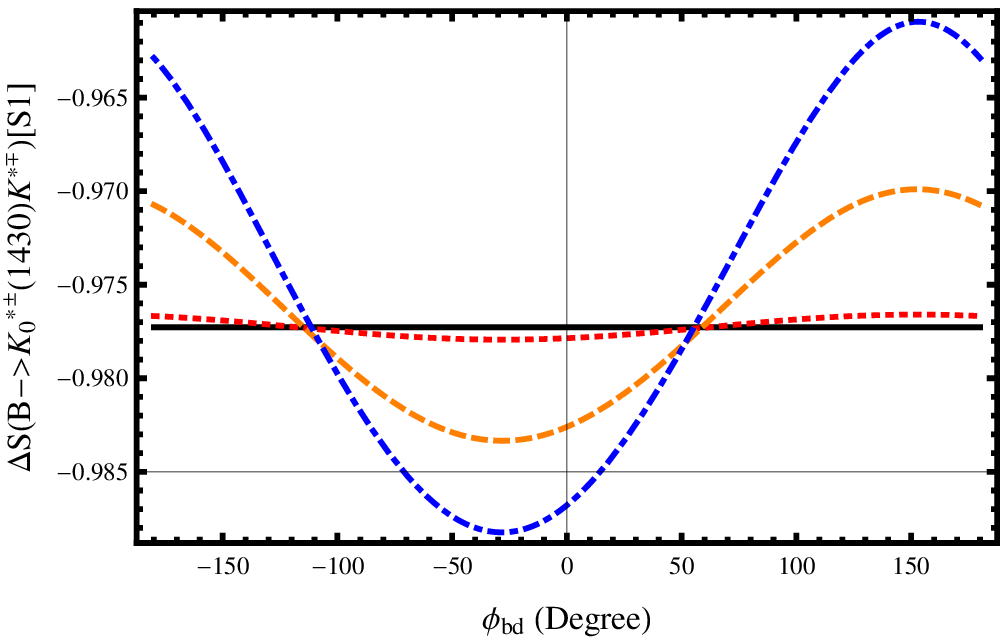}\hspace{2cm}
\includegraphics[scale=0.65]{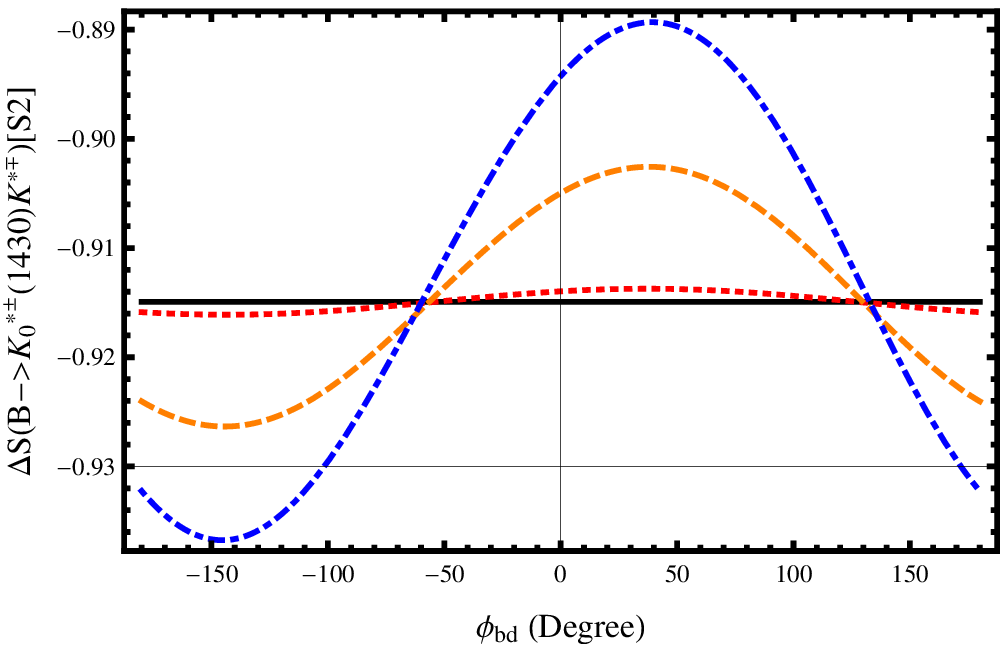}
\caption{Under different scenarios (S1 and S2), the CP-violating parameters $A_{CP}$, $C$, $\Delta C$, $S$  and $\Delta S$  ($\%$) of $B \to K_0^*(1430)K^{*}$ as  functions of the weak phase $\phi_{bd}$, the dotted (red), dashed (orange) and dot-dashed (blue) lines represent results from the $\zeta=0.001,~0.01,~0.02$, and the solid lines (black) are the predictions of SM.} \label{Fig:cpsv}
 \end{center}
 \end{figure}

The relations of the direct $CP$ asymmetries $A_{\mathrm{CP}}^{\mathrm{dir}}$, $C$, $\Delta C$, $S$ and $\Delta S$ with $\phi_{bd}$ with different $\zeta$ are also presented in Fig.\ref{Fig:cpsp} and Fig.\ref{Fig:cpsv}, for $B \to  K_0^{*\pm}(1430)K^{\mp}$ and $B \to  K_0^{*\pm}(1430)K^{*\mp}$, respectively. From Fig.\ref{Table:cp}, the observables of $CP$ asymmetries are much sensitive to the  $Z^\prime$ than the branching fractions. For example, under S1, the direct asymmetry of $\overline B^0\to K_0^{*\pm}(1430)K^{\mp}$ is $79\%$ in SM, while it would change to $49\%$ with $Z^\prime$ boson. In future, these observables could be used to probe the effect of new physics. If the $Z^\prime$ were  detected in the colliders directly, these decays would also be useful to constrain the couplings.

\section{Summary}

Within the QCD factorization  approach, we have studied the pure annihilation type decays $B \to K_0^{*\pm}(1430)K^{(*)\mp}$ in SM and the family non-universal leptophobic $Z^\prime$ model. Both the branching fractions and the $CP$ asymmetry observables have been calculated. The branching fractions we predicted are 1-2 orders of magnitude smaller than the results from PQCD approach. When the $Z^\prime$ involved, if $m_{Z^\prime}>1\mathrm{TeV}$ ($\zeta <0.01$), its contributions will be buried by the large uncertainties of SM. If $m_{Z^\prime}\approx 600\mathrm{GeV}$ ($\zeta=0.02$), both the branching fractions and $CP$ asymmetries of $\overline B^0\to K_0^{*+}(1430)K^-$ could be changed remarkably, which provides us a place for probing the effect of new physics. These results are hopeful to be tested in  Belle-II, LHC-b or the future   high energy circular  colliders.

\end{multicols}
\begin{multicols}{2}

\end{multicols}
\end{document}